\documentclass[twocolumn,showpacs,preprintnumbers,amsmath,amssymb,superscriptaddress,prb,floatfix]{revtex4}

\usepackage{graphicx}
\usepackage{dcolumn}
\usepackage{bm}
\usepackage{longtable}

\begin{document}

\setlength{\parskip}{0 pt}

\preprint{APS/123-QED}

\bibliographystyle{unsrt}

\title{A high resolution, hard x-ray photoemission investigation of La$_{2-2x}$Sr$_{1+2x}$Mn$_2$O$_7$ ($0.30\leq x<0.50$): on microscopic phase separation and the surface electronic structure of a bilayered CMR manganite}

\author{S. de Jong}
\email{sdejong@science.uva.nl}
\affiliation{Van der Waals-Zeeman Institute, University of Amsterdam, NL-1018XE
Amsterdam, The Netherlands}
\author{F. Massee}
\affiliation{Van der Waals-Zeeman Institute, University of Amsterdam, NL-1018XE Amsterdam, The Netherlands}
\author{Y. Huang}
\affiliation{Van der Waals-Zeeman Institute, University of Amsterdam, NL-1018XE Amsterdam, The Netherlands}
\author{M. Gorgoi}
\affiliation{Helmholtz Zentrum Berlin GmbH, Albert-Einstein-Strasse 15, 12489 Berlin, Germany}
\author{F. Schaefers}
\affiliation{Helmholtz Zentrum Berlin GmbH, Albert-Einstein-Strasse 15, 12489 Berlin, Germany}
\author{J. Fink}
\affiliation{Helmholtz Zentrum Berlin GmbH, Albert-Einstein-Strasse 15, 12489 Berlin, Germany}
\author{A. T. Boothroyd}
\affiliation{Clarendon Laboratory, Oxford University, Oxford, OX1 3PU, United Kingdom}
\author{D. Prabhakaran}
\affiliation{Clarendon Laboratory, Oxford University, Oxford, OX1 3PU, United Kingdom}
\author{J. B. Goedkoop}
\affiliation{Van der Waals-Zeeman Institute, University of Amsterdam, NL-1018XE
Amsterdam, The Netherlands}
\author{M. S. Golden}
\affiliation{Van der Waals-Zeeman Institute, University of Amsterdam, NL-1018XE
Amsterdam, The Netherlands}

\date{\today}

\begin{abstract}
Photoemission data taken with hard x-ray radiation on cleaved single crystals of the bilayered, colossal magnetoresistant manganite La$_{2-2x}$Sr$_{1+2x}$Mn$_2$O$_7$ (LSMO) with $0.30\leq x<0.50$ are presented. Making use of the increased bulk-sensitivity upon hard x-ray excitation it is shown that the core level footprint of the electronic structure of the LSMO cleavage surface is identical to that of the bulk. Furthermore, by comparing the core level shift of the different elements as a function of doping level $x$, it is shown that microscopic phase separation is unlikely to occur for this particular manganite well above the Curie temperature.
\end{abstract}


\maketitle

\begin{figure} [!tb]
\begin{center}
\includegraphics[width=1.0\columnwidth]{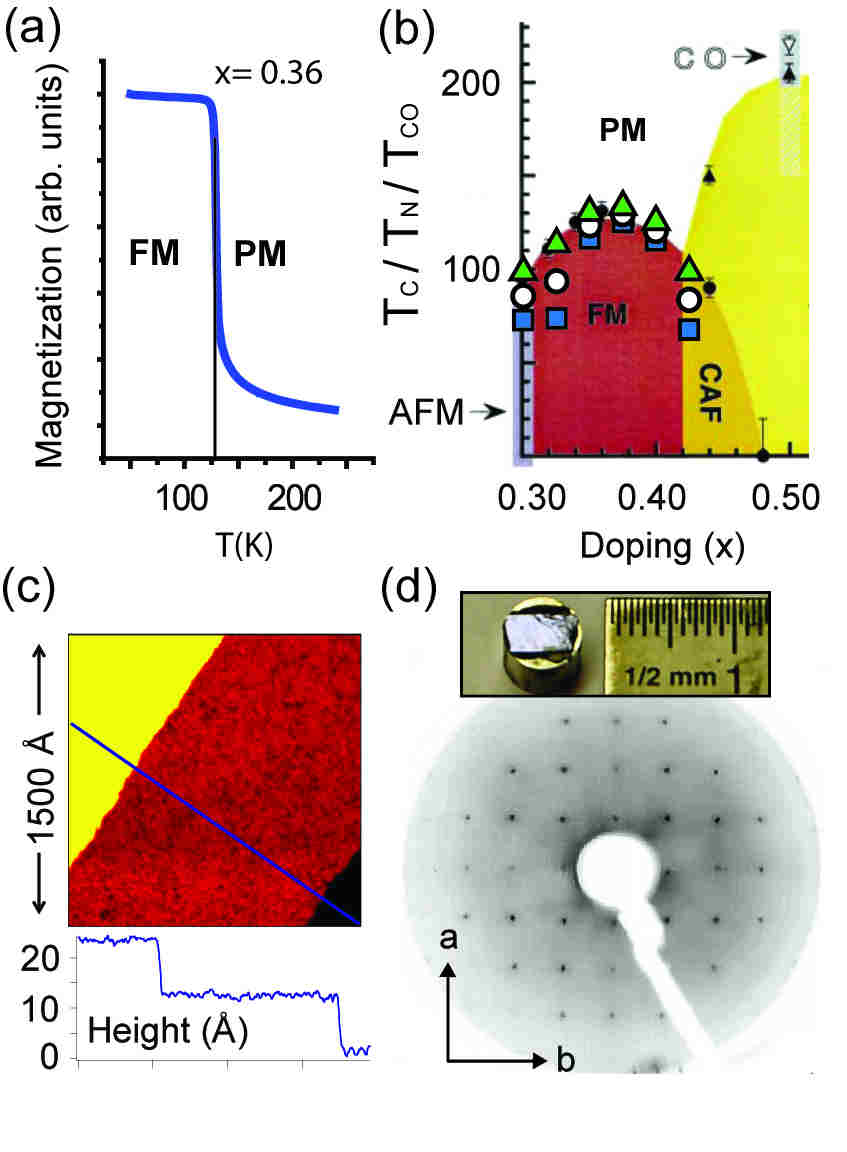}
\caption{\label{fig:pic1}Sample and cleavage surface quality of the LSMO crystals. (\textbf{a}) Magnetization versus temperature for an $x=0.36$ sample measured using SQUID magnetometry. Data taken after zero field cooling, with an external field $B=100$~G $\parallel c$. The sample shows a sharp transition from a paramagnetic (PM) to a ferromagnetic (FM) state at $T_{C}=130~K$ with a total width smaller than 5~K, (\textbf{b}) Magnetic transitions as measured using SQUID magnetometry of LSMO with $x=0.30$ ($T_N$) and $0.325\geq x \geq 0.425$ ($T_{C}$), plotted on top of the magnetic phase diagram, taken from Ref. [\onlinecite{Ling}]. Depicted are the onset, midpoint and endpoint temperatures of the transition (triangles, white circles and squares), (\textbf{c}) STM topograph (150$\times$150~nm$^2$) of an in vacuum cleaved LSMO crystal ($x=0.35$), taken at $T=4$~K showing three flat atomic terraces. The blue line indicates the trace of the line scan depicted below the topograph. The step heights correspond to half the $c$-axis length of the tetragonal unit cell (11 \AA). The terraces themselves are very smooth, with a height corrugation of the order of only 1\AA\ over tens of nanometers, (\textbf{d}) LEED image of a typical LSMO sample, $E=400$~eV showing a very clear tetragonal pattern, without any signs of a structural reconstruction. The inset shows a cleaved crystal on top of a cleavage post, with a smooth and mirror-like surface over millimeters. (color online)}
\end{center}
\end{figure}

\begin{figure*} [!tb]
\begin{center}
\includegraphics[width=2.0\columnwidth]{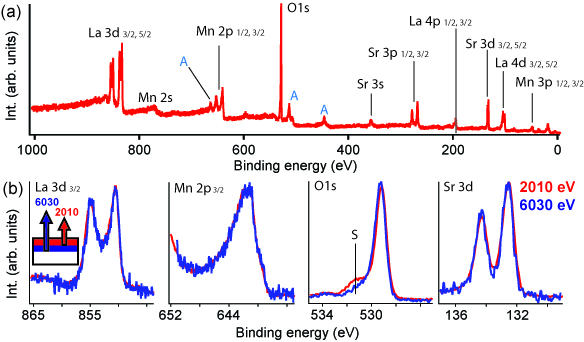}
\caption{\label{fig:pic2}(Hard) x-ray photoemission data on LSMO ($x=0.30$). (\textbf{a}) Overview core level spectrum taken with $h\nu=2160$~eV at room temperature with the main core levels labeled. Auger peaks are indicated with an `A', (\textbf{b}) Zooms of representative core levels for all four elements of LSMO: La 3d$_{3/2}$, Mn 2p$_{3/2}$, O 1s and Sr 3d$_{3/2, 5/2}$ taken with $h\nu=2010$ and 6030~eV. Only the O 1s core level peak shows a small surface contribution, either from adsorbed residual gas on the sample surface or from the sample holder, indicated with `S' in the O 1s spectrum. The inset in the La 3d spectrum illustrates schematically the increased bulk sensitivity of the data recorded using 6030~eV radiation compared to $h\nu=2010$~eV. (color online)}
\end{center}
\end{figure*}

\begin{figure} [!tb]
\begin{center}
\includegraphics[width=1.0\columnwidth]{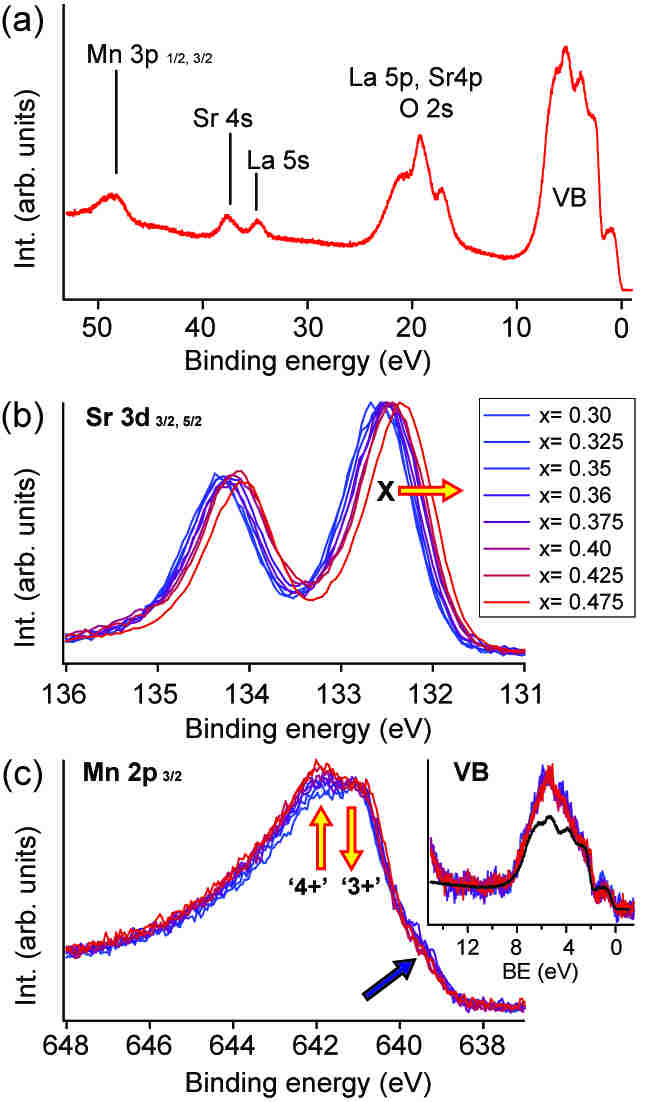}
\caption{\label{fig:pic3} Photoemission data on LSMO. (\textbf{a}) Angle integrated photoemission spectrum of the near- valence band region, taken with $h\nu=140$~eV at $T=20$~K ($x=0.375$), (\textbf{b}) The Sr 3d core level spectra for all measured doping levels taken with $h\nu=2010$~eV at room temperature, normalized to their maximum intensity, (\textbf{c}) The Mn 2p$_{3/2}$ core level spectra for all measured doping levels taken with $h\nu=2010$~eV at room temperature. The blue arrow indicates the non-locally screened feature. The yellow arrows indicate locally screened features for Mn$^{4+}$ and Mn$^{3+}$. The spectra are normalized to the Mn$^{3+}$ feature at 641~eV. The inset shows the valence band spectra for the various doping levels recorded with $h\nu=2010$~eV and in black the valence band recorded with $h\nu=140$~eV for comparison. The difference between the valence band data recorded with 2010~eV and 140~eV is due to the different O 2p and Mn 3d photoionization cross-sections for the two excitation energies. (color online)}
\end{center}
\end{figure}


\section*{Introduction}
After the high temperature superconductors, one of the most studied correlated electron systems in condensed matter physics are the colossal magnetoresistant (CMR) manganites. These systems show an insulator-to-metal transition on cooling, that coincides with the onset of long range ferromagnetic order. This transition goes paired with colossal changes in the magnetoresistance. \cite{CMR_first, Tokura_overview} The ability to alter the electronic properties of these materials by applying a magnetic field makes them very interesting for applications. From a more fundamental point of view though, despite years of research, the microscopic origin of the colossal magnetoresistance effect in these systems is still the subject of much debate.

Early theoretical attempts to explain the (colossal) magnetoresistance displayed by the manganites focused on the double exchange mechanism where the ferromagnetism facilitates metallicity in a strongly Hund's rule coupled system, by increasing the hopping parameter of an $e_g$ electron that is aligned parallel to the $t_{2g}$ electrons of the neighboring manganese sites, compared to anti-parallel or randomly aligned spins. \cite{Kubo} The double exchange mechanism however, can only account for a change of resistivity of about 30 percent across the ferro-to-paramagnetic transition, while some of the CMR manganites, such as La$_{2-2x}$Sr$_{1+2x}$Mn$_2$O$_7$ (abbreviated forthwith LSMO, with $x$ being the hole doping, i.e. increasing the Mn$^{4+}$ to Mn$^{3+}$ ratio), display changes that are a factor 100 larger. \cite{Millis, Moritomo}

One of the main contemporary groups of models attempting to explain the CMR effect is based on an electronic phase separation scenario, that focuses on the idea that with hole doping, instead of a continuous change, the density of the Mn e$_{g}$ electrons is unstable for certain doping concentrations leading to a spatial separation of charge into patches of higher and lower than nominal hole doping, while the distribution of dopant atoms would be homogeneous in the sample. \cite{Yunoki} The propensity toward phase separation is expected to be especially high near half-doping, where most manganites show anti-ferromagnetic, orbital and charge order due to the charge disproportionation into equal amounts of formally Mn$^{4+}$ and Mn$^{3+}$ ions.

Much of the experimental support for phase separation comes from studies involving surface sensitive techniques such as scanning tunneling microscopy/ spectroscopy (STM/ STS) \cite{Fath, Becker, Zhang, Roessler} or (angle resolved) photoemission [(AR)PES]. \cite{Sarma, Ebata, Dessau_NaturePhys} As both techniques have direct access to the electronic structure of a material they would be well suited to address the issue of CMR, as the root of this phenomenon clearly lies in the electronic nature of the manganites. Moreover, STM has the advantage to be a spatially resolving technique on the micro- to nanometer scale, and would thus be extremely useful in (dis-)proving a phase separation scenario. Yet, the reported length scales at which the phase separation would be evident range from micro- to nanometers and often the reported phase separation has little correspondence with the magnetic transition temperature. So in many cases the observed phase separation could well be caused by for instance sample inhomogeneity or (for thin films substrate induced) lattice strain, rather than by \textit{electronic} phase separation. 

From the (AR)PES side, in particular focusing on (bilayered) LSMO, some studies carried out in the metallic part of the phase diagram report a pseudo-gapped Fermi surface \cite{Dessau_science} while others have reported the existence of small quasiparticle-like peaks at the Fermi level, \cite{Dessau_NaturePhys, Shen, Dessau, UsPRB, Mannella, Sun2} followed by a large incoherent spectral weight at higher binding energies. One would expect the temperature dependence of these quasiparticle peaks, which normally are associated with the metallic phase, to track the bulk Curie temperature, but it does so only in some datasets. \cite{Mannella} In other studies the sharp quasiparticle-like feature persists up to temperatures of order two times $T_C$, thus well into the insulating regime. \cite{Dessau_NaturePhys, UsPRB} Furthermore, an x-ray resonant magnetic scattering/ STS study on air-cleaved bilayered LSMO single crystals found the first bilayer at the surface of this material to be insulating and magnetically unordered at low temperatures, in contrast to what was found for the bulk. \cite{XRMS} In this light an STM/ STS study on the anti-ferromagnetic $x=0.30$ compound is also worth mentioning, where the tunneling spectrum of the entire probed surface (over many thousands of measurements) of an in vacuum cleaved sample seemed to be gapped (hence insulating), while bulk- resistivity measurements again showed a metallic characteristic. \cite{Renner}

All in all, the picture arising from the mentioned STM and ARPES studies is rather diffuse, sometimes even inconsistent and often in contrast with the physical properties measured by bulk probes such as resistivity and magnetization versus temperature. An important question is therefore whether the surface electronic structure of bilayered LSMO is indeed identical to the bulk one. The cleavage plane of this compound is generally assumed to be in between two rock-salt layers. The (La,Sr)O surface termination layer thus obtained is stoichiometrically identical to those in the bulk of the crystal, but not charge neutral and could consequently be reconstructed electronically in order to avoid a polarization catastrophe. 

In this paper, we present a doping dependent hard x-ray photoemission study on bilayered LSMO, La$_{2-2x}$Sr$_{1+2x}$Mn$_2$O$_7$, with $0.30\leq x \leq 0.475$. Although several x-ray photoemission studies on (perovskite) LSMO exist in the literature, for example Refs. [\onlinecite{Matsuno}] and [\onlinecite{Offi}], the majority has been conducted either on polycrystals or on single crystals cleaved in air or poor vacuum, disqualifying a comparison between surface and bulk electronic properties. This study is conducted on properly in vacuum cleaved single crystals and carried out using excitation radiation in the hard and soft x-ray regime on a wide range of doping levels across the metallic part of the phase diagram. Making use of the increased bulk sensitivity with higher excitation energies (several nanometers for 6~keV radiation, instead of a typical $\approx1$~nm for VUV excited ARPES experiments and Al K$\alpha$ x-ray photoemission studies), owing to the increased mean free path length of escaping photoelectrons with higher kinetic energy, we show that the surface electronic structure of bilayered LSMO is identical to that of the bulk. Furthermore, evaluating the core level shift per element as a function of doping, we show that the chemical potential of bilayered LSMO is not pinned upon approaching half doping, putting strong constraints on the temperature range and length scales at which phase separation could occur for these samples.

\section*{Experimental}
Experiments with photon energies around $h\nu=2$ and $6$~keV were performed at the double crystal monochromator KMC-1 beamline at Helmholtz Zentrum Berlin, Berlin, coupled to the \textit{Scienta} R4000 analyzer of the HiKE endstation \cite{HIKE}. Experiments were carried out at room temperature in a grazing incidence geometry with a total energy resolution of 300 and 180~meV for $h\nu=2$ and $6$~keV, respectively, as determined from the width of the Fermi edge of a piece of gold foil. Single crystals of LSMO were grown using the traveling floating zone technique in Amsterdam ($x=0.30$, 0.36 and 0.40) and in Oxford ($x=0.30$, 0.325, 0.35, 0.375, 0.40, 0.425 and 0.475). \cite{ox_growth} The quality of the crystals, the Curie temperature and the sharpness of the metal-to-insulator transition were checked by magnetometry measurements using a superconducting quantum interference device (SQUID), see Fig. \ref{fig:pic1}a and b. Prior to the photoemission measurements, the single crystals were cleaved at room temperature in a vacuum better than $1\times10^{-9}$~mbar, resulting in shiny, flat cleavage surfaces (see the inset to Fig. \ref{fig:pic1}d). The results obtained from crystals from both Amsterdam and Oxford with the same nominal doping level were identical.

The experiments on the core level shift versus doping were repeated with a lab-based Al~K$\alpha$ source from \textit{VG-Scienta} coupled to a \textit{Specs} PHOIBOS 100 hemisperical analyzer with a total energy resolution of $\approx1$~eV. These LSMO single crystals, from the same batches as the crystals used for the hard x-ray experiments, were cleaved at room temperature in a vacuum better than $5\times10^{-10}$~mbar. The results obtained from the lab-system (though not shown in this paper) were identical to the results obtained with hard x-ray radiation at the synchrotron. 

Referencing of the binding energy scale was done by measuring the kinetic energy of the 4f core levels of a gold film in electrical contact with the measured cleavage surface. The accuracy of the binding energy referencing was better than 50~meV and 100~meV for the experiments with $h\nu=2010$ and $6000$~eV, respectively (whereby this error is dominated by the drift in temperature of the monochromator crystals as the current in the storage ring decreases with time). Furthermore, it was tested that the samples were not charging electrically (either due to the contact resistance of the sample mounting or an intrinsically low conductivity of the samples), by varying the incident photon flux and confirming that the measured kinetic energy of the core level peaks remained unchanged.

The cleavage surfaces obtained from the in vacuum cleaved single crystals were of excellent quality, as shown by STM measurements and low energy electron diffraction on similarly prepared LSMO crystals, yielding topographs with very clean, flat surfaces and perfectly tetragonal diffraction patterns without any signs of reconstructions whatsoever, see Fig. \ref{fig:pic1}c and d. 

\section*{Results}
In Fig. \ref{fig:pic2}a an overview spectrum of LSMO taken with $h\nu=2160$~eV is shown, displaying many identifiable core level lines. Zooms of representative core levels for all four elements taken with $h\nu=2010$ and 6030~eV are depicted in  panel (b) of Fig. \ref{fig:pic2}. Comparing the more bulk sensitive 6030~eV with the more surface sensitive 2010~eV data, it is immediately clear that the binding energies are almost identical and that the line shapes exactly match. Only the O 1s spectrum shows a small surface related shoulder at the high binding energy (BE) side of the main line, but it is likely that this feature either comes from a small amount of residual gas adsorbed on the sample surface, or from the sample holder and mount.\cite{footnote-grazing} For all measured crystals and doping levels, the 6030~eV and 2010~eV data were similar to a very high degree to that shown in Fig. \ref{fig:pic2}b.

In Fig. \ref{fig:pic3}a an angle integrated photoemission spectrum taken with an excitation energy of 140~eV, displaying the valence band and several shallow core levels is shown. At these photon energies the escape depth of the excited photoelectrons is minimal and will hardly exceed the first unit cell of bilayered LSMO. Still, the shallow core levels depicted do not show any sign of significant shoulders or a peak form deviating from a simple, single Gaussian-broadened Lorentzian, thus providing more evidence that the surface and bulk electronic structure are indistinguishable.

Figure \ref{fig:pic3}b shows the Sr 3d core level spectra ($h\nu=2010$~eV) for all measured doping levels. Clearly, the core level spectra shift (almost monotonically) toward lower binding energy with increasing hole doping, but their peak form is basically unaltered. The Sr 3d spectra are also representative for the doping dependent behavior displayed by the La and O core levels, all of which shift rigidly toward lower BE without changing peak form as the doping level of the single crystals is increased. This is in contrast with the Mn core levels. Figure \ref{fig:pic3}c shows that the Mn core levels remain essentially unshifted with increased doping, but they do change their peak form slightly. The rather broad Mn 2p$_{3/2}$ peak consists of several components, with a feature at BE$\approx639$~eV associated with a non-locally screened core hole \cite{non_local} (see blue arrow in Fig. \ref{fig:pic3}c) and around 642~eV two distinguishable locally screened features associated with Mn$^{4+}$ (higher BE) and Mn$^{3+}$ (lower BE), whose relative intensity alters qualitatively in line with the expected trend for increased hole doping (indicated with yellow arrows in Fig. \ref{fig:pic3}c). The inset to Fig. \ref{fig:pic3}c shows the Mn 3d dominated valence band spectra of all doping levels recorded with $h\nu=2010eV$. Similar to the Mn 2p core levels, the depicted valence bands do not notably shift over the entire doping series. Also obvious changes in their line shapes do not occur.

\begin{figure}
\begin{center}
\includegraphics[width=1.0\columnwidth]{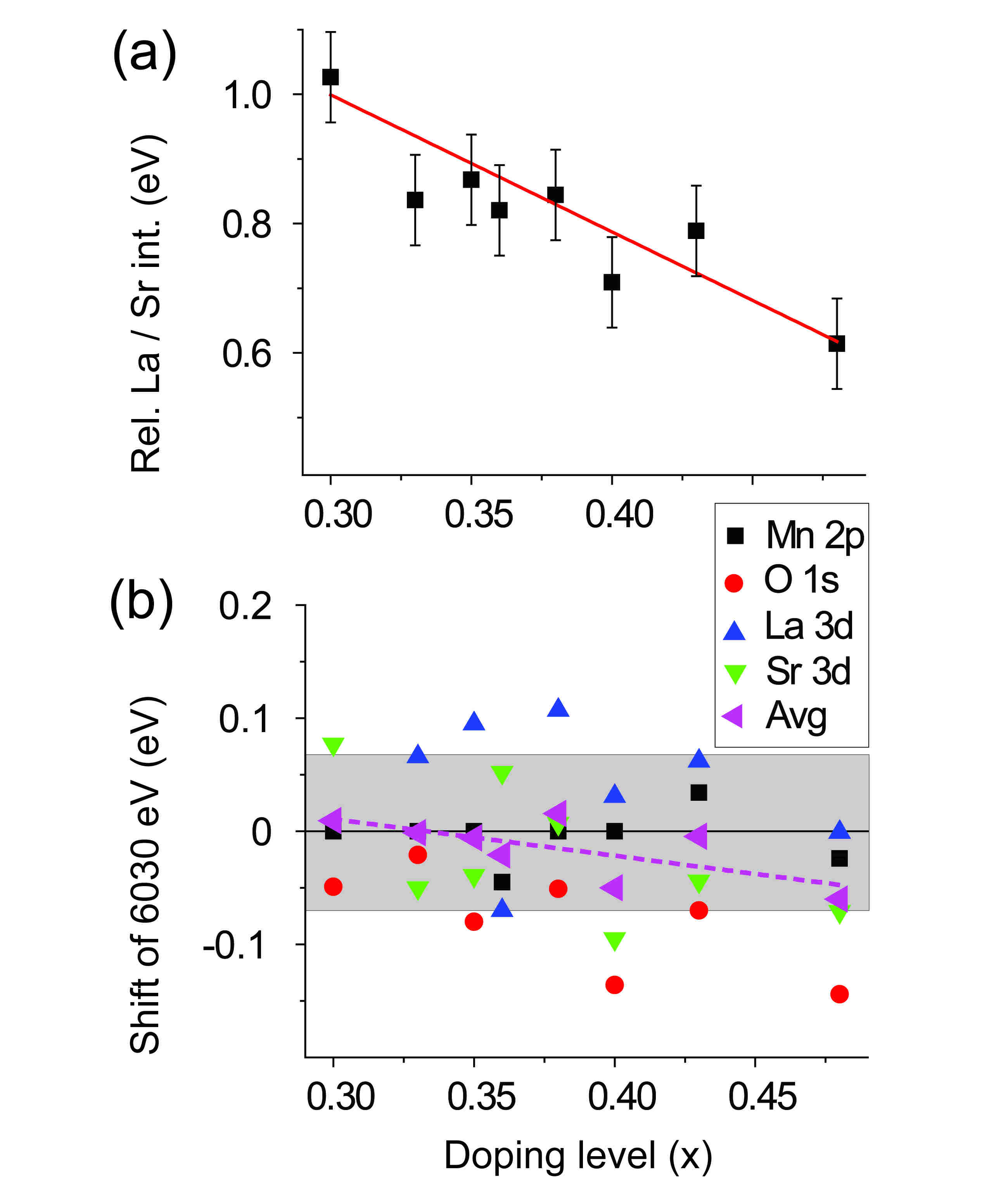}
\caption{\label{fig:pic4}Doping dependence of the core level data taken with $h\nu=2010$~eV at room temperature. (\textbf{a}) The relative intensity ratio of the La 4d and Sr 3d core level peaks. The red line is the expected ratio from the stoichiometry of the nominal compositions. For details on the normalization, see footnote [\onlinecite{norm}]. The error bars for the measured intensity ratio are caused by uncertainty in the background subtraction procedure and in the correction of the decaying beam intensity versus time, (\textbf{b}) Shift in binding energy of the core levels measured with $h\nu=6030$ compared to the 2010~eV data for all doping levels and all four elements (a positive shift means one toward lower binding energy). The pink triangles are the averaged value per doping level over all 4 core levels. The pink dotted line is a linear fit to these averaged values. The error margin in the determination of the relative shift is indicated with the gray shaded area. The individual error bars for the data points in panel (b) have been omitted for clarity. (color online)}
\end{center}
\end{figure}

\section*{Discussion}
From the characterization of the bilayered LSMO crystals by measurement of their bulk magnetization versus temperature, Fig. \ref{fig:pic1}b, it is clear that the magnetic properties of the samples are in line with what is expected from literature for their nominal doping. \cite{Ling} One extra check to ensure that the measured cleavage surfaces are representative for their expected hole doping, is to compare the evolution of the ratio between the La and Sr core level photoemission peaks throughout the doping series, as this quantity should change in a predictable manner. The straight red line in \ref{fig:pic4}a shows the stoichiometric La to Sr ratio versus doping, normalized to that of $x=0.30$. The black symbols are the measured and normalized \cite{norm} intensity ratios between the La 4d and Sr 3d core levels. The error in the measured values is due to a variation in the background intensity from sample to sample and the fact that the flux from the beamline changed over time, due to the decaying ring current of the synchrotron. Despite these uncertainties, the measured La/Sr ratios agree (within the error bars) with the expected doping dependence, indicating that the measured cleavage surfaces are indeed representative for their nominal doping level.

Now let us look a little bit closer at the surface versus bulk issue as regards the electronic structure of LSMO. As mentioned in the previous section, both photoemission spectra taken with VUV and hard x-ray radiation do not resolve surface related features for the core levels, strongly suggesting the existence of a similar surface and bulk electronic structure for bilayered LSMO. There is, however, a small but detectable shift of binding energy between the data taken with $h\nu=2010$~eV and 6030~eV. In Fig. \ref{fig:pic4}b this shift is plotted for all four elements and for all measured doping levels. The error bars are mainly determined by the binding energy referencing of the $h\nu=6030$~eV data, which could be executed with an accuracy of about 100~meV. As can be seen from the gray shaded band in panel (b), the majority of the data points fall within the error bars. Taking the averages of the shift for all elements per doping level, only a very weak downward trend with increasing hole doping is visible, with a change of shift that is only 50~meV over the entire doping series. We can therefore safely conclude that also the apparent shift between the $h\nu=2010$~eV and 6030~eV data is not indicating a systematic, significant difference between the bulk and surface of bilayered LSMO in terms of hole doping level or charge transfer.

\begin{figure} [!tb]
\begin{center}
\includegraphics[width=1.0\columnwidth]{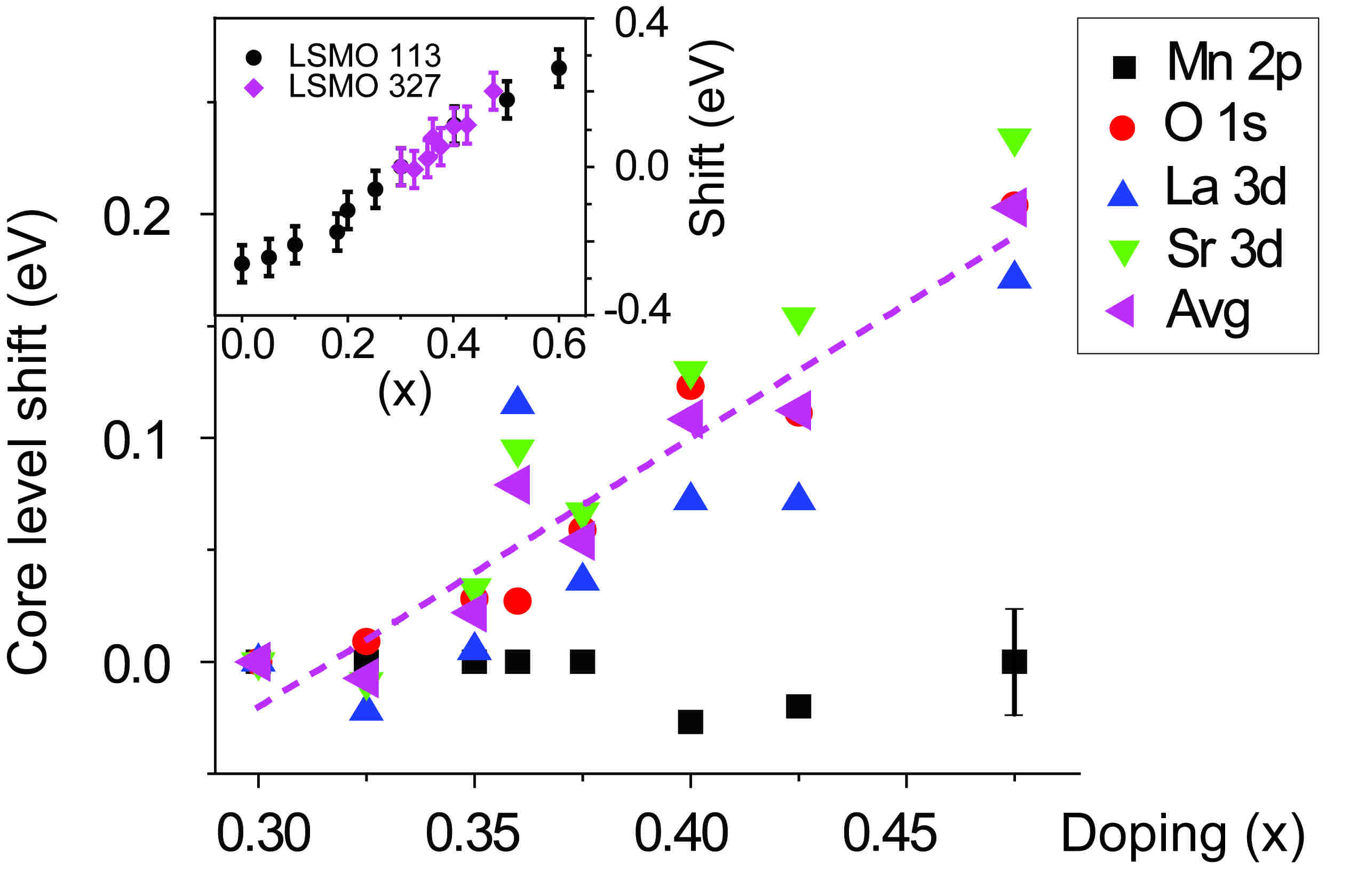}
\caption{\label{fig:pic5}Shift in binding energy versus doping with respect to $x=0.30$ for the La 3d, Sr 3d, Mn 2p and O 1s core levels ($h\nu=2010$~eV, room temperature). Pink symbols are the averaged shift per doping level of La 3d, Sr 3d and O 1s. The pink dotted line is a linear fit to these averaged values. The error bars for all data points are represented by the error bar depicted for the $x=0.475$ Mn 2p data point: the rest of the error bars have been omitted for clarity. The inset shows the chemical potential shift with respect to $x=0.30$ for perovskite LSMO (LSMO 113) after Ref. [\onlinecite{Matsuno}] compared to the measured chemical potential shift for bilayered LSMO (LSMO 327, pink symbols). Note that in contrast to the bilayered case, doping levels below $x=0.30$ can be obtained for the perovskite. (color online)}
\end{center}
\end{figure}

In Fig. \ref{fig:pic5} the results for the shift in binding energy per element as a function of doping are summarized. The shifts are plotted relative to the binding energy measured for $x=0.30$, thereby setting the shift for this doping level to zero. The shift values were determined by cross-correlating the core level spectra of one particular element for the different doping levels with each other, which is justified as the form of the core levels hardly changes with doping. \cite{shift_comment} As mentioned (and also shown in Fig. \ref{fig:pic3}), all core levels shift toward lower binding energy with increasing doping (a positive shift), except for the Mn peaks, that stay constant in BE. Taking the averages of the La, Sr and O shifts, a linear trend is evident for bilayered LSMO, as shown by the pink dotted line in Fig. \ref{fig:pic5}. These results fall exactly in line with the doping dependence of the binding energy for the perovskite analogue (La,Sr)MnO$_3$ \cite{Matsuno}, see the inset to Fig. \ref{fig:pic5}.  As also argued in Ref. [\onlinecite{Matsuno}], in general, the shift of a core level $\Delta E$ can be explained by a number of terms given by the following formula:
\\
\\
$\Delta E = \Delta \mu + K \Delta Q + \Delta V_{M} -\Delta E_{R}$,
\\
\\
where $\Delta \mu$ is the change in the chemical potential, $K\Delta Q$ the change in the number of valence electrons of the atom under consideration (the chemical shift), $\Delta V_{M}$ the change in the Madelung potential, and $\Delta E_{R}$ is the change in the extra-atomic relaxation energy due to polarizability of the atoms and the conduction electrons surrounding the created core hole. \cite{hufner} A significant contribution of the Madelung potential to the shift in binding energy seen in Fig. \ref{fig:pic5} can be excluded since the O 1s and the Sr 3d and La 4d shift in the same direction and the Madelung term has opposite sign for anions and cations. Moreover, the contribution due to changes in the screening of the core-hole potential by polarizable surrounding ions and metallic-conduction electrons ($\Delta E_{R}$) can be discarded as the measurements are performed at room temperature, i.e in the insulating regime, and, additionally, the polarizabilty is not expected to change with Sr doping, as La$^{3+}$ and Sr$^{2+}$ have a very similar cation radius (and thus polarizability, which is proportional to the size of the atom). As also the number of valence electrons of La, Sr and O are not expected to change across the doping range, the observed linear trend in the binding energy shift for these three elements reflects the shift of the chemical potential with hole doping. The fact that the Mn core levels show no shift may be due to the fact that with doping the Mn valency does change, whereby the chemical shift counterbalances the shift of the chemical potential.

This result directly gives important insight into the applicability of phase separation scenarios to the bilayered manganites. Several numerical studies \cite{Yunoki, Moreo} have shown that phase separation in manganites in the clean limit would lead to a pinning of the chemical potential as a function of doping. Chemical potential pinning has indeed been shown for the perovskite manganite (Pr,Ca)MnO$_3$, that has a narrow bandwidth compared to LSMO and a larger propensity toward charge and orbital order. \cite{Ebata} From simulations it has been predicted that disorder in the lattice (by for instance cation substitution) can lift the pinning of the chemical potential and the size of the phase separated clusters, at which this pinning would occur, increases. \cite{Moreo2} The chemical disorder in LSMO however is low, owing to the similar cation radius of La$^{3+}$ and Sr$^{2+}$, and the fact the chemical pinning does occur in the case of the more cation radius mismatched (Pr,Ca)MnO$_3$ compound, supports the idea that phase separation does not occur for bilayered LSMO in the measured temperature and doping regime.

Thus, the monotonic shift of the chemical potential clearly observed in Fig. \ref{fig:pic5} renders a scenario involving phase separation into (large, $>$nm) clusters highly unlikely. The fact that this outcome is not different to that reached for cubic LSMO, indicates that fluctuations (which will be greater in the 2D system) may well lead to a reduction in the Curie temperature (maximally 130~K for bilayered LSMO, but well above room temperature for the cubic case), but not to a different doping dependent behavior of the chemical potential. Also the proximity of the charge and orbitally ordered phase at $x=0.475$ (that is within a few percent of the half doped case) does not seem to provoke macroscopic phase separation in the LSMO case.

At this stage it is worth mentioning that preliminary experiments carried out on a lab-system using Al~K$\alpha$ radiation at liquid nitrogen temperatures (so well below the magnetic transition temperature of the entire doping range studied) showed a chemical shift versus doping behavior that is similar to the one measured in the hard x-ray experiments at room temperature, thus suggesting that the absence of electronically phase separated clusters not only holds for temperatures far into the paramagnetic region of the phase diagram, but also in the low temperature, ferromagnetic part.

To summarize, by carrying out photoemission experiments using different excitation energies in the hard x-ray regime we have shown that the surface electronic structure of the bilayered CMR manganite La$_{2-2x}$Sr$_{1+2x}$Mn$_{2}$O$_{7}$, $0.30\leq x\leq0.475$, as probed by core level spectroscopy, is identical to the bulk electronic structure. By examining the binding energy shift of the core levels as a function of doping, we can show that the chemical potential shifts monotonically as a function of hole doping, without any sign of pinning over the investigated doping range. Therefore it is highly unlikely that phase separation occurs over macroscopic ($>$nm) length scales for bilayered LSMO at temperatures well away from the magnetic and metal-to-insulator transition.
\\
\\
We thank Jesse Klei for valuable experimental input and Huib Luigjes for expert
technical support. This work is part of the research programme of the `Stichting voor Fundamenteel Onderzoek der Materie (FOM)', which is financially supported by the `Nederlandse Organisatie voor Wetenschappelijk Onderzoek (NWO)'. We also acknowledge funding from the EU (via I3 contract RII3-CT-2004-506008 at Helmholtz Zentrum Berlin).


\begin{thebibliography}{}
\bibitem{footnote-grazing} Due to the extreme grazing incidence geometry necessary to better match the inelastic mean-free path length of the photoelectrons and the penetration depth of the hard x-rays, it is unavoidable to also pick up signals from the sample mounting.
\bibitem{norm} The measured intensities of core levels for the different elements are determined not only by their relative stoichiometry, but also by their (photon energy dependent) photoionisation cross-sections. To allow a comparison between the measured La/ Sr ratio and the ratios expected from the nominal doping levels, the measured La 4d to Sr 3d peak area ratio versus doping has been scaled such that the average measured value over all doping levels coincides with the average ratio predicted from stoichiometry over the entire doping range. Thereafter, all data in the plot has been scaled such that the predicted ratio for $x=0.30$ is set to 1. This way the reported, experimentally determined ratios are independent of the cross-sections for the La 4d and Sr 3d core levels.
\bibitem{shift_comment} It was checked and confirmed that binding energy shifts versus doping, obtained by either a Voigt-peak fitting procedure or a center of gravity determination of the core level spectra, yielded the same result as the cross-correlation procedure.
\end{thebibliography}
\end{document}